A molecular perspective on the yield and flow of polymer glasses: The role of enhanced segmental dynamics during active deformation

*M.D. Ediger and Kelly Hebert, Department of Chemistry, University of Wisconsin – Madison, Madison, WI 53706*

The mechanical properties of polymer glasses are often critical in determining the best material for a particular application. Extremely stiff materials (high modulus) may be important for some applications while avoiding catastrophic failure due to fracture (high toughness) may be more important for others. The mechanical properties of a polymer glass will depend upon both molecular structure and many experimental variables, including temperature and the mode of deformation (tension, compression, or shear).

In this chapter we discuss the mechanical response of polymer glasses from a molecular perspective. In particular, we consider how deformation changes the rate at which polymer segments rearrange and how this in turn influences the mechanical response of the material. It will be shown that this focus on the changes in *dynamics* provides an understanding of many important features of polymer glass deformation. Of course, it is also true that the *structure* of a polymer glass must be altered by nonlinear deformation. Although not a major focus, we will make some comments about these structural changes at the end of this chapter.

*12.1 Introduction to polymer glass deformation*

In order to start our discussion, let's consider one type of deformation experiment used to characterize polymer glasses. Figure 1 shows "stress-strain" curves that describe the response of a polymer glass to a constant strain rate deformation. For these data, the sample was deformed uniaxially in tension although qualitatively similar data are also obtained in compression. These particular experiments were performed at $T_g$ – 19 K on a lightly cross-linked poly(methyl methacrylate) (PMMA) glass after annealing the sample at the testing temperature for about 30 minutes; here $T_g$ denotes the glass transition temperature obtained with differential scanning calorimetry at heating rate of 10 K/min. These details are specified for completeness; qualitatively similar data could be obtained for almost any polymer glass (including uncross-linked samples) at a similar temperature below $T_g$. The different curves shown in Figure 1 come from experiments performed at different strain rates as indicated.

Figure 1 is useful for identifying several characteristic features of polymer glass deformation. At extremely small strains (perhaps up to 0.002), the stress increases linearly with strain. If the stress is released after such a deformation, the sample returns very nearly to its original length; these small deformations are essentially reversible. The slope of the stress-strain curve at such small deformations defines the Young's modulus E. The modulus is essentially independent of strain for such small strains and this mechanical experiment can be considered to be in the linear response regime. Larger deformations are non-reversible and are not in the linear response regime. As shown in Figure 1, a maximum in the stress is reached (the yield stress) at a strain of about 0.03. Larger strains result in a decrease of the stress (strain softening). At even larger strains than those shown in the figure, the stress increases again (strain hardening). The area underneath the stress-strain curve up to the point where the sample breaks is a measure of toughness. As discussed below, strain hardening plays a critical role in delocalizing the deformation response of a polymer glass; such delocalization is necessary for a tough material.

Figure 1 also shows the influence of strain rate on polymer glass deformation and this provides an interesting method to quantify the nonlinearity of these experiments. A useful comparison can be made with the predictions of the Maxwell model. The Maxwell model is the simplest model for the mechanical response of a viscoelastic liquid; it can be represented by a purely elastic spring (with modulus E) in series with a purely viscous dashpot (with viscosity $\eta$). Even though the Maxwell model has only a single relaxation time ($\tau = \eta/E$), it is a useful starting point for understanding linear viscoelastic behavior. Using this model, calculations for three constant strain rate experiments are displayed in Figure 2, with the range of strain rates matching the range in Figure 1. The Maxwell model predicts that the stress will initially rise linearly with strain and then reach a steady-state value (the flow stress $\sigma_{flow}$), with $\sigma_{flow}$ depending linearly upon the strain rate (a factor of 6 for the strain rate range displayed). In contrast, note that the stresses at a strain of 0.10 in Figure 1 differ by only 10%. The near-independence of the post-yield stress data with respect to strain rate, in contrast to the prediction of linear viscoelasticity, indicates that these deformations are deeply in a nonlinear regime.

*12.2 The role of enhanced segmental dynamics during deformation*

We can use the Maxwell model to provide further insight into why the response of a polymer glass to deformation is so highly nonlinear. The Maxwell model predicts the following value for the flow stress: $\sigma_{flow} = E\tau\dot{\varepsilon}$. Here $\tau$ is the relaxation time, which we identify as the time for segmental rearrangements of the polymer, and $\dot{\varepsilon}$ is the strain rate. For PMMA at $T_g - 19$ K, we can estimate that $\tau = 50000$ s in the absence of deformation [1]. Using a strain rate of $\dot{\varepsilon} = 3.1 \times 10^{-5}$ s$^{-1}$ and the measured value of E, we can calculate from the Maxwell model that $\sigma_{flow}$ should be about 2500 MPa, a value about 150 times larger than the flow stress shown in Figure 1. If we turn this calculation around and use the observed flow stress during deformation as an input, we can calculate that the *effective* segmental relaxation time during deformation is ~350 s, or about 150 times shorter than the segmental relaxation time in the absence of deformation. As we will show below, this estimate is in semiquantitative agreement with experiments that directly measure segmental dynamics during deformation.

The calculation in the previous paragraph gives us a reasonable qualitative picture of the deformation of polymer glasses: something about the deformation causes the segmental dynamics to speed up enormously and this allows the polymer glass to flow at much lower stresses than otherwise would have been possible. This is actually an old idea that goes back at least as far as Eyring. Eyring [2] predicted that stress accelerates the rate at which barriers will be overcome in a solid. This can be interpreted as "landscape tilting". The multidimensional potential energy landscape (PEL) governs the thermodynamics and dynamics of a system at constant volume. At any given instant, the configuration of a system is specified by its position on the PEL. Eyring calculated that stress decreases the heights of some barriers on the PEL and increases the heights of others. The system evolves in the direction of the decreased barriers and thus stress accelerates dynamics.

The idea that deformation enhances the rate of segmental dynamics has been included in models of polymer glass deformation for the last 50 years. Some models follow Eyring [2] in using stress [3-10] or strain rate [11,12] as the control variable. Others use strain [9], configurational entropy [13,14], configurational internal energy [15,16], the amplitude of density fluctuations [17-19], or free volume [20,21]. Many of these models make use of a *material* or *effective time* formalism [8,22,23]. In this approach, a linear mechanical model is used to predict the nonlinear deformation of a polymer glass with one modification: the relaxation time specifying the polymer segmental dynamics is allowed to change during the deformation in response to stress or some other control variable. As a preview, we warn the reader that experiments show that no simple mechanical variable (such as stress or strain rate) can predict how segmental relaxation times evolve during all phases of deformation.

While the above paragraphs have emphasized the nonlinearity of the mechanical response as a key feature of polymer glass deformation, we wish to briefly mention other aspects of this problem that are particularly challenging. *Polymer glasses are out of equilibrium* from a thermodynamic perspective and thus these systems are constantly evolving due to structural relaxation (physical aging) [24-26]. Structural relaxation certainly influences the material response during and immediately following deformation as well. In addition, it is known that *dynamics in polymer glasses are spatially heterogeneous*, i.e., segments in one region of the sample can relax on time scales orders of magnitude faster than segments in another region only a few nanometers away [27-29]. There is no reason that these regions of differing dynamics should respond uniformly to deformation. The experiments shown below indicate that there are substantial changes in the heterogeneous dynamics during deformation; slower regions experience dynamics that are enhanced by a greater extent than are the faster regions.

Several of the ideas presented above in Sections 12.1 and 12.2 will be revisited in the remainder of this chapter. Section 12.3 will discuss how segmental dynamics are monitored during deformation, and will also introduce the optical probe technique featured in this chapter. The behavior of segmental mobility during constant strain rate and constant stress mechanical protocols, as measured using the optical probe technique, will be illustrated in Sections 12.4 and 12.7, respectively. Section 12.5 includes a discussion of experiments which compare mechanical and optical measures of segmental dynamics. The effect of temperature on segmental dynamics during deformation is discussed in Section 12.6. Comments about changes in structure during deformation will be made in Section 12.8.

*12.3 Experimental methods for measuring segmental dynamics during deformation*

Prior to work in our lab, which began about seven years ago, there were experimental results showing indications that segmental dynamics were enhanced during the deformation of a polymer glass. Solid-state NMR experiments were used to study deformation-induced segmental mobility in the amorphous regions of a semicrystalline nylon 6 sample [30]. Other experiments demonstrated that the diffusion of a plasticizer into poly(ether imide) glass under compression far below $T_g$ was similar to diffusion at $T_g$ in the absence of deformation [31]. While both of these approaches showed enhanced dynamics as a result of deformation, neither quantified the change in the average segmental relaxation time during active deformation.

The optical technique [1,32-39] developed in our lab to monitor segmental dynamics during deformation relies on the observation that the reorientation of a molecular probe can be a good reporter of segmental dynamics. We typically utilize the fluorescent probe DPPC (N,N'-Dipentyl-3,4,9,10-perylenedicarboximide, shown in Figure 4), which is dissolved in the polymer matrix at a concentration of ~$10^{-6}$ M. In polymer melts above $T_g$, this probe and other similar probes have ensemble-average reorientation times that closely track the polymer segmental relaxation times [32]; the probe reorientation time is typically longer than dielectric measurements of the segmental relaxation process, but the ratio of these two times is essentially independent of temperature. There is also good evidence that the probes reasonably track segmental dynamics in the glassy state during physical aging [37]. We assume that the close correspondence between probe reorientation and polymer segmental dynamics is also valid during deformation and the results below support this view.

To measure probe reorientation and mechanical deformation at the same time, we need optical access to the polymer glass sample. Figure 3 shows how this is accomplished.

Panel a provides the sample dimensions; the sample is thin enough to ensure that deformation-induced birefringence does not interfere with the optical polarization measurements described below. Panel b shows a deformation apparatus that fits on top of a confocal optical microscope [1]. A computer controls the linear actuator and reads the load cell, allowing experiments at constant strain rate, constant stress, or combinations of these experiments with stress or strain relaxation. The temperature-controlled cell that houses the polymer glass sample has an optical window on the bottom to allow optical access prior to, during, and after deformation.

Probe reorientation is measured with an optical photobleaching method. At the start of the experiment, probe molecules in the glass are orientationally isotropic. A linearly polarized laser excites a subset of probe molecules and a small fraction of these are permanently photobleached; the photobleached molecules are not orientationally isotropic but preferentially have orientations that allow them to efficiently absorb the polarized laser excitation. This step, which requires a fraction of a second, establishes an anisotropic distribution of *unbleached* probe molecules. As probe molecules reorient, this anisotropic orientational distribution will evolve into an isotropic distribution. This evolution is monitored by illuminating the sample with a weak reading beam of circularly polarized light. At the start of a measurement, the fluorescence that results from the reading beam displays the highest intensity in the polarization orthogonal to the excitation beam polarization. The unbleached probes attain an isotropic distribution on a timescale that defines the probe relaxation time $\tau_{probe}$; as a result, the fluorescence from the reading beam becomes unpolarized. From these experiments, we obtain a time-dependent anisotropy decay function r(t) that describes the reorientation of the ensemble of probe molecules.

Figure 4 shows that segmental dynamics are enhanced during deformation. The figure shows anisotropy decay functions for DPPC in lightly cross-linked PMMA obtained just prior to deformation (in black) and during various stages of a constant strain rate deformation experiment; this particular data was acquired in the same experiment as the green curve shown in Figure 1. The more rapid decay curves obtained during deformation indicate that the ensemble of probes reorients more rapidly during deformation and thus segmental relaxation is also occurring more rapidly. Each of these anisotropy decay curves is fit to the Kohlrausch-Williams-Watts function (KWW):

$$r(t) = r(0)e^{-(t/\tau_{\mathrm{probe}})^{\beta}}$$

Here $\tau_{probe}$ is a characteristic reorientation time, β is a parameter that characterizes the shape of the decay process ($\beta < 1$ is a stretched exponential decay), and r(0) is the initial anisotropy. Another characteristic reorientation time discussed in this chapter is $\tau_c$, which is obtained by integrating r(t)/r(0). Although $\tau_c$ and $\tau_{probe}$ both track the segmental dynamics of the polymer, $\tau_{probe}$ emphasizes a faster portion of the relaxing segments as compared to $\tau_c$ [34].

In order to interpret these results, we need to track the evolution of the local strain during deformation. Because these samples neck during extension and do not deform homogeneously, the local strain is not equal to the global strain (which is controlled by the linear actuator). Local strain is determined by tracking the evolution of the distance between lines that are photobleached on the sample within a few hundred microns of the spot where mobility is measured. Prior to yield, the local strain very nearly equals the global strain. After yield, the local strain generally exceeds the global strain by a factor of 2-3 (because we perform our experiments close to the point where necking originates) [1].

Before discussing our results in more detail, we briefly describe one other approach to measuring segmental dynamics during deformation. Subsequent to our development of the optical method described above, Lesser and coworkers [40] have developed a dielectric technique for measuring segmental relaxation during deformation. Electrodes are brought into contact with the two opposite sides of the polymer glass sample to form the capacitor needed for this method. The results obtained indicate that dynamics are enhanced during deformation and qualitatively match the optical experiments in other aspects as well. While this approach has the advantage that the polymer segments are being monitored directly (rather than probe molecules), the extent to which the average segmental dynamics are changed during deformation has not yet been quantified with this method.

*12.4 Segmental mobility during constant strain rate deformation*

From data such as those shown in Figure 4, we can track the evolution of the average segmental relaxation time during a constant strain rate deformation experiment. Figure 5 shows these changes for experiments at $T_g$ – 19 K at five different strain rates; the data set indicated by circles corresponds to the experiment shown in Figure 4. Each experiment shows a significant shortening of the probe reorientation time as yield is approached followed by a post-yield regime where the probe time is almost constant. At the lowest strain rates, the mobility increases about a factor of 25 during deformation while at the highest strain rates, mobility increases of more than a factor of 100 are observed. Figure 6 shows another important feature of these experiments; the KWW $\beta$ parameter increases during deformation, indicating a narrowing of the distribution of relaxation times. A change of the KWW $\beta$ parameter from 0.3 to 0.6 is very substantial, indicating that the distribution of relaxation times narrows from more than 3 decades to about 1 decade [41]. We note that the KWW $\beta$ parameter is sensitive to the size of the probe molecule used in these experiments. While the qualitative trends shown in Figure 6 are expected to apply to the distribution of segmental relaxation times, the absolute values of the KWW $\beta$ parameter may be shifted to some extent [34].

Prior to detailed analysis of this data, we return briefly to the Maxwell model described earlier. We noted that if we use the Maxwell model to calculate an effective relaxation time during deformation, that relaxation time would need to be shortened by more than a factor of 100 in response to deformation. The data in Figure 5 shows changes of this magnitude. This indicates that the qualitative interpretation of the stress/strain experiment in terms of enhanced segmental dynamics is consistent with our direct measurements of probe relaxation times during deformation.

The qualitative features of Figure 5 are precisely those expected in light of recent computer simulations. Riggleman, de Pablo, and coworkers used molecular dynamics computer simulations of coarse-grained polymer chains to study deformation in the glassy state [42]. Their results show a strong decrease (up to a factor of 1000) in segmental relaxation times prior to yield followed by a nearly constant post-yield relaxation time; the simulations indicate that the post-yield relaxation time depends strongly on strain rate. While these simulations did not detect large changes in the KWW $\beta$ parameter during deformation, they did detect dynamics that were more spatially homogeneous in the flow regime [38]. This greater spatial uniformity is consistent with the narrowing of the distribution of relaxation times observed in Figure 6 and provides a useful spatial interpretation of these results. Recent simulations by Rottler and coworkers [43,44] are also consistent with these experimental results.

A particularly useful feature of the simulations by Riggleman and de Pablo [42] is that they were performed both in tension and compression. One possible interpretation of the data in Figure 5 is that segmental mobility increases with strain because free volume increases; it is known that the density of glassy polymers decreases during extension. The simulations observed essentially identical values of the segmental relaxation time during tension and compression as a function of strain, as long as the strain rates were the same. This result argues against a free volume interpretation of the segmental relaxation time changes that occur in a stress/strain experiment, since the simulated polymer glass increased in volume during extension but densified during compression [45]. This result is particularly valuable since it has not yet been possible to perform optical measurements of probe mobility in polymer glasses during compression.

The features in Figure 5 also match predictions based upon a molecular theory of Chen and Schweizer [17]. In response to a constant strain rate deformation, they predict that the segmental relaxation time shortens dramatically in the pre-yield regime but remains essentially constant post-yield. They further predict that at constant temperature, the post-yield value of the segmental relaxation time scales with strain rate to the power -0.86. Figure 7 shows a plot of our experimental data in this format. The data obtained at $T_g$ – 11 K shows a slope of -0.80 ± 0.06, which is compatible with the theory. A more detailed comparison can be found in reference 39. The theory of Chen and Schweizer does not consider spatially heterogeneous dynamics or a distribution of relaxation times, and thus their approach cannot account for the results shown in Figure 6. A recent mesoscopic constitutive approach of Medvedev and Caruthers [16] does predict changes in the distribution of relaxation times and is qualitatively consistent with the results shown in Figure 6.

We now return to the question of what controls the segmental relaxation time during deformation. Clearly the answer cannot simply be *strain rate*, as can be seen in the pre-yield regime of Figure 5; here the strain rate is constant while the relaxation time is changing significantly. The answer cannot simply be *stress* either, for the relaxation time is roughly constant in the post-yield regime while the stress is decreasing (see Figure 1). The theory of Chen and Schweizer [17] allows a molecular interpretation of the results shown in Figure 5. In this theory, the segmental relaxation time is controlled by two factors. Stress decreases the relaxation time through an Eyring-like mechanism (landscape tilting). In addition, the amplitude of local density fluctuations influences the relaxation time; we interpret this variable to indicate the position of the system on the PEL. The Chen/Schweizer approach indicates that landscape tilting is responsible for the pre-yield decrease in the relaxation time; below we show experimental data consistent with this view. In the post-yield regime, the two mechanisms operate together to maintain a constant relaxation time; as the deformation progresses into the strain softening regime, the decrease in the stress means that landscape tilting contributes less to the enhanced mobility but the system is being pulled up higher on the PEL (where energy barriers are lower) and this effect increases mobility enough to make up the difference.

*12.5 Can segmental mobility during deformation be measured from a purely mechanical experiment?*

The experiments described above make a compelling case that changes in segmental dynamics are intimately connected with the nonlinear mechanical deformation of polymer glasses. As such, it is reasonable to ask if there isn't some other (easier!) way to obtain these segmental relaxation times than the optical measurements described above. For example, if these relaxation times could be obtained from a purely mechanical measurement, this would allow much broader access to this fundamentally important information. There is a significant history of efforts along these lines, including the work of Yee [46], Martinez-Vega [47], and their coworkers. There are also cautionary notes in the literature about the difficulty of obtaining molecular relaxation times during a nonlinear mechanical measurement [48]. Inspired by recent work from Caruthers,

Medvedev, and coworkers [49-51], we have tested one particular idea for obtaining segmental relaxation times from a purely mechanical measurement. Our approach was to perform the optical measurement of probe reorientation during the proposed mechanical measurement so that the relaxation time derived from mechanical measurements could be directly compared to the probe reorientation time. As we describe below, we found only a partial correspondence between the mechanical and probe relaxation times.

Figure 8 shows mechanical measurements performed on a lightly cross-linked PMMA glass in which a constant strain rate deformation is followed by a stress relaxation experiment (during stress relaxation, the strain is fixed and decay of stress is measured as a function of time). The four experiments shown in the figure all share a common strain rate and differ only in the strain at which stress relaxation was initiated. It has been proposed that the initial rate of stress relaxation is an accurate measurement of the rate of segmental dynamics [50,51]. Figure 9 shows an overlay of the stress relaxation experiments in a format that normalizes the data to the initial stress and shifts time to overlap the starting times of the stress relaxation portion of the experiments. Following references 50 and 51, we define a mechanical relaxation time from the inverse of the initial decay rates of these curves (determined from a linear fit); thus $\tau_{mech}$ is about a factor of two shorter at a strain of 0.098 than at a strain of 0.003.

Figure 10 compares the mechanical relaxation times extracted as described in the previous paragraph to optical measurements of probe reorientation that were obtained during the same experiments. For simplicity, we present only the probe measurements obtained during the constant strain rate portion of the experiment; we make this comparison because $\tau_{mech}$ was proposed as measure of mobility during constant strain rate deformation. As expected, the probe relaxation times have behavior consistent with Figure 5; $\tau_{probe}$ changes by about a factor of 30 in the pre-yield regime. In contrast, $\tau_{mech}$ changes by only a factor of 2 in the pre-yield regime. In the post-yield regime, $\tau_{mech}$ is similar to $\tau_{probe}$; we have preliminary results that confirm that this agreement in the post-yield regime is also obtained at other strain rates.

Unfortunately, Figure 10 indicates that we have not yet succeeded in our effort to identify a purely mechanical experiment that tracks the segmental dynamics throughout all stages of a constant strain rate deformation. Our understanding is that $\tau_{probe}$ accurately tracks the average segmental dynamics. In contrast, $\tau_{mech}$ is also influenced by the width of the distribution of relaxation times. In a linear viscoelastic experiment, $\tau_{mech}$ would increase as a function of strain merely because there is a distribution of relaxation times, even though these relaxation times would not be changing during the deformation. Because the actual mechanical experiment shown is highly nonlinear, we speculate that the values of $\tau_{mech}$ shown in Figure 10 are the combination of two effects. As the strain increases, the distribution of relaxation times tends to increase $\tau_{mech}$ while the nonlinear deformation tends to decrease $\tau_{mech}$. Apparently, these two effects that influence $\tau_{mech}$ nearly cancel under these conditions. It remains to be established if any purely mechanical experiment can directly track changes of segmental relaxation times during all phases of nonlinear mechanical deformation of a polymer glass.

Although $\tau_{mech}$ does not provide the same information as $\tau_{probe}$ and the KWW $\beta$ parameter, it joins these parameters as observables that characterize (in various ways) the influence of a constant strain rate deformation on the dynamics of polymer glasses. We anticipate that these three observables in concert provide quite a demanding test of theories, models, and simulations.

*12.6 Effect of temperature on segmental dynamics during deformation*

Temperature has a very strong influence on the deformation properties of polymer glasses. Taking polycarbonate as an example, we note that the yield stress doubles from 30 MPa to 60 MPa as the temperature is dropped from $T_g$ – 20 K to $T_g$ – 100 K (for this comparison, we use a series of experiments in extension with the strain rate of $4 \times 10^{-3}$ $s^{-1}$ [52]). We can think about this increase in yield stress with decreasing temperature in qualitative terms as follows. At a lower temperature, the timescale for segmental relaxation of a given glass will be longer, since less thermal energy is available to surmount energy barriers in the PEL. On the other hand, at yield, we anticipate that the segmental relaxation time will be approximately the same no matter what the deformation temperature, since the segmental dynamics need to be fast enough to enable flow at the strain rate imposed by the experiment (see next paragraph for more details on this point). Thus we anticipate that, at low temperature, the segmental relaxation time must change by a much larger factor in the pre-yield regime than at high temperature. According to the work of Chen and Schweizer [17], the landscape tilting mechanism dominates in the pre-yield regime, so higher levels of stress will be required at low temperature in order to drive the system to yield, in general agreement with the experimental data discussed above.

We have performed a series of experiments to quantitatively test the effect of temperature on the segmental relaxation time during constant strain rate deformation [39]. The results are qualitatively consistent with the scenario from the previous paragraph but differ in one important detail. The post-yield segmental dynamics of the same polymer glass deformed at the same strain rate, but at different temperatures, are somewhat faster at higher temperatures. This effect is shown in Figure 7 for lightly cross-linked glasses of PMMA. We interpret the faster dynamics at higher temperature in Figure 7, even for the same strain rate, to mean that thermally-activated barrier crossing is quite important in the post-yield regime for these polymer glasses. Under the conditions of our experiments, we estimate [39] that the free energy barriers that are crossed in the flow state are still very substantial (with barrier heights of about 39 kT during flow vs. 45 kT in the absence of deformation).

Our conclusion that thermal barrier crossing is very important for segmental dynamics during post-yield deformation is somewhat different than recent theoretical and simulation work. As discussed in reference 39, the temperature dependence shown in Figure 7 is larger than anticipated by the theory of Chen and Schweizer. The role of thermal barrier crossing has also been investigated in computer simulations by Chung and Lacks [53,54]; these authors came to the conclusion that barrier hopping plays a relatively minor role during deformation. There is certainly an opportunity for additional computer simulations on this topic, particularly those which calculate the instantaneous segmental relaxation time during deformation across a wide range of temperature.

*12.7 Segmental mobility during creep deformation*

There are two reasons for investigating changes in segmental relaxation times during different types of deformations. First, a wide variety of deformation schemes are utilized to characterize polymeric materials; measurements of molecular motion during these experiments will be useful for understanding the observed mechanical response. Second, different deformation schemes provide an opportunity to check our understanding of the fundamental factors that control changes in segmental dynamics. In creep deformation, stress is held constant and this provides a way to test the extent to which stress controls mobility during deformation.

Figure 11 shows results for two creep experiments on lightly cross-linked PMMA samples [37]. In a creep experiment, the stress is held constant while the strain is the dependent variable. Panel a shows the changes in the local strain during these creep experiments; the increase in the strain corresponds to the creep portion of the measurement while the subsequent decrease occurs after the stress is set to zero (i.e., strain recovery). Both of these experiments show a large permanent set, as indicated by the persistent strain even after long times in the absence of stress. Panel b shows the changes in segmental relaxation times during each of these experiments. One important feature of these experiments is that these samples were aged to equilibrium prior to deformation; the experiment temperature was sufficiently close to $T_g$ that equilibrium could be achieved in one day of aging. Thus, in a technical sense, these deformation experiments started in the equilibrium supercooled liquid state. The equilibrium relaxation time at the temperature of the mechanical experiment is marked in panel b. It is important to note the segmental relaxation time has nearly returned to its predeformation (equilibrium) value by the end of the experiment; in contrast, the strain remains at a large value. This result indicates that segmental dynamics are not directly controlled by strain. In panel c, the changes in the KWW β parameter during these experiments are presented. The β parameter is highly correlated with instantaneous segmental relaxation times such that large β parameters are observed when the relaxation times are the shortest. We interpret this to indicate that the distribution of segmental relaxation times narrows as the average segmental relaxation time decreases.

The behavior shown in Figure 11 is consistent with the interpretation presented earlier for constant strain rate experiments. In order to connect the behavior shown in Figure 11 to the constant strain rate experiments, we note that almost all the creep data shown in this figure was obtained in the flow regime. In a creep experiment the onset of flow is analogous to yield in a constant strain rate experiment. All the data shown for the flow regime in Figure 11 demonstrate the same correlation between strain rate and segmental relaxation time as was indicated for the constant strain rate data in Figure 7. That is, when the strain is increasing most rapidly (the highest strain rate), the segmental relaxation time is the shortest. The KWW β parameter behavior shown in Figure 11c is consistent with the behavior shown for constant strain rate experiments in Figure 6; in the post-yield flow regime, larger values of β are associated with higher strain rates. The qualitative features shown in panels a and b of Figure 11 have also been observed in molecular dynamics computer simulations of polymer glasses subjected to creep and strain relaxation [35,38].

For the purposes of this chapter, the most insightful creep deformation experiments are those in which the stress is too low to cause flow. In such an experiment, the strain changes very little and the relaxation time can be measured very accurately at constant engineering stress. The results of many such experiments are shown in Figure 12 [34]. These experiments were performed at three temperatures below $T_g$ for lightly cross-linked PMMA glasses. The three dashed lines are fits to the Eyring model $\tau \propto \sigma/\sinh\left(\frac{\sigma \cdot V}{2 \cdot k_b T}\right)$ where $k_bT$ is the thermal energy, with one fitting parameter (the activation volume, V). For each temperature, the Eyring model accurately describes the effect of stress on the segmental relaxation time up to true stress levels of about 10 MPa. As the samples enter the flow regime at higher true stresses, deviations from the Eyring behavior are observed.

The results shown in Figure 12 are completely consistent with our earlier interpretation of constant strain rate experiments. In the pre-yield regime of both creep and constant strain rate experiments, the landscape tilting mechanism described by the Eyring equation accurately describes the observed segmental relaxation times. Once flow occurs, segmental relaxation during creep is faster than would be expected on the basis of landscape tilting alone. During flow, the system is being pulled up the PEL into a regime in which energy barriers are lower; this second mechanism is then responsible for the deviations from Eyring behavior shown in Figure 12. Chen and Schweizer [18] have developed a theory for polymer glass deformation during creep and our experiments [32,33,36,37] are qualitatively consistent with the predictions of their theory. The interpretation provided in this paragraph is also consistent with their theory, if we identify the larger local density fluctuations discussed in the theory with higher regions of the PEL.

*12.8 Changes in polymer structure during deformation*

Structure/property relationships are at the heart of materials science and it is useful to address the role of structure changes in the deformation of polymer glasses.

When a macroscopic polymer glass sample is deformed by 20% or 50%, it is clear that there is a significant change in the structure of individual chains. One manifestation of this is the birefringence developed by polymer glasses during deformation (and typically maintained after deformation ceases). The observed value of the birefringence depends upon both local and large-scale rearrangements of the polymer chains and this effect has been explored extensively by Osaki, Inoue, and their coworkers [55,56]. Neutron scattering experiments allow the changes in structure caused by deformation to be determined more directly. A recent experiment compared deformed and undeformed polymer glasses, and concluded that chains are deformed affinely on the scale of ~10 nm (roughly $R_g$ for the chains studied) but are isotropically arranged on a much smaller length scale (~2 nm) [57]. At a qualitative level we understand this result to mean that the segmental mobility during deformation is sufficient to extensively rearrange segments on a small length scale; on the length scale of the entire polymer chain, however, polymer chains do not have sufficient mobility to relax. Molecular dynamics computer simulations on non-polymeric glassy systems (i.e., spherical particles) also indicate that large-scale deformation is affine while more local rearrangements are not [58].

Changes in polymer structure on large length scales are highly relevant for an understanding of the strain hardening that occurs at large deformations. Strain hardening determines the extent to which strain is localized during extension, with higher levels of strain hardening favoring delocalization of the strain, which in turn leads to higher ductility [59]. Strain hardening cannot occur if the molecular weight of the polymer chains is too low or if the entanglement density of the system is too low, and under these conditions polymer glasses fail via fracture at low strains [60,61].

We take the point of view that the yielding of polymer glasses (up to strains at which strain hardening remains insignificant) is a generic glass problem that can be understood without considering polymer molecular weight or large-scale changes in polymer structure. Yielding phenomena do not depend significantly upon polymer molecular weight as long as the molecular weight is high enough to ensure that yield occurs before failure. Consistent with this view, we have preliminary results indicating that glasses of uncross-linked polymer chains give rise to the same results as those shown for lightly cross-linked samples in Figures 5 and 7, if the temperature is adjusted to constant $T - T_g$. Furthermore, we expect that the detailed molecular structure of different polymers (polystyrene, polycarbonate, or PMMA) will not have a significant influence on the main features shown in Figures 5-7. We also have preliminary results that support this statement.

The very large changes in dynamics that occur during deformation in our experiments are ultimately connected to changes in the structure of the glass at a very local level. In the theory of Chen and Schweizer [17], the segmental relaxation time is controlled by two mechanisms. Prior to yield, the landscape tilting mechanism dominates; we interpret this to mean that very small changes in local structure, induced by the applied stress, lower some energy barriers and allow faster dynamics. At later stages of the deformation, the amplitude of local density fluctuations also influences the time scale of segmental relaxation in the glass. The amplitude of these density fluctuations increases in a typical deformation of a polymer glass and this is one of the factors that enhance dynamics in the post-yield regime of a constant strain rate experiment (see Figure 2 of Reference 16). In the context of the Chen and Schweizer theory, these local density fluctuations are the coarse-grained representation of the local structural changes that occur in the post-yield regime of the deformation of a polymer glass. As Chen and Schweizer have noted, the predicted changes in the amplitude of these local density fluctuations is quite small and only a few attempts have been made to measure these changes with x-ray scattering [62]. We should not be surprised that a very small change in structure can give rise to a large change in dynamics as this is a generic feature of the glass problem. For example, it is also difficult to pinpoint the changes in local structure that are responsible for the super-Arrhenius temperature dependence of many supercooled liquids.

As the yielding of polymer glasses appears to be part of a more generic yielding problem that includes other classes of amorphous materials, we wish to briefly comment on this connection. At a phenomenological level, the yielding of metallic and colloidal glasses is similar to that of polymer glasses [63,64]. Much of the literature on the yielding of metallic and colloidal glasses focuses on shear transformation zones, which are local rearrangements in structure that occur in response to deformation. A useful feature of colloidal glasses is that direct imaging of these local rearrangements is possible. For example, Schall et al. [65] showed that these rearrangements have the geometric features expected for shear transformation zones. In their experiments on colloids, local structural rearrangements were shown to occur via thermal barrier crossing, with barriers that are almost as high during deformation as they would be in the absence of deformation; our experiments on polymer glasses discussed above agree with their conclusions [39]. If the shear transformation zone models and the model of Chen and Schweizer are both correct, then we expect that shear transformation zones are the local rearrangements that are represented in a coarse-grained manner by the local density fluctuations of Chen and Schweizer. Of course, there are also important differences between colloidal glasses and polymer glasses. One manifestation of this is the difference in the characteristic barrier height during deformation near $T_g$ (~18 kT for colloids [65] vs. ~39 kT for polymer glasses [39]). In spite of these differences, we expect that studies of the deformation of other classes of amorphous materials will continue to deepen our understanding of polymer glass deformation.

### *12.9 Concluding remarks*

The deformation of polymer glasses is an important and challenging problem that is not yet sufficiently understood. In industry, the deformation properties of polymer glasses are generally predicted using models which must be extensively parameterized against experimental data. Because these models are not based upon a completely accurate fundamental understanding of the deformation process, such models often fail when applied outside the range of experiments from which they were parameterized; that is, a model that accurately predicts the results of a constant strain rate deformation may fail completely in describing a cyclic deformation. Our experiments, in combination with simulations and more fundamental theory/modeling, have the goal of advancing our understanding to the point where the right physics can be included in models used to predict the properties of polymer glasses in an industrial setting. One could be optimistic that such a model could accurately predict the mechanical response of a polymer glass to many different types of deformations, over time periods encompassing the 40-year lifespan of a product.

Our experiments indicate that enhanced segmental dynamics is a key feature of the deformation of polymer glasses. In future experiments, we hope to more critically examine the underlying mechanisms of enhanced segmental dynamics in theories. For example, the work of Chen and Schweizer indicates that both landscape tilting and the increased amplitude of density fluctuations contribute to enhanced mobility during a constant strain rate experiment. We expect that experiments which reverse the application of stress or strain during deformation can sensitively determine the separate contributions of each mechanism (since a tilted landscape can be un-tilted by removing stress).

Our experiments also indicate that changes in the width of the distribution of segmental relaxation times during deformation can be very substantial. It is unclear what role these changes play in the macroscopic mechanical properties of polymer glasses. At present, the Chen and Schweizer approach cannot describe this effect, although the theory predicts several features which are qualitatively consistent with our results. The model of Caruthers and Medvedev demonstrates changes in the width of segmental relaxation times and provides at least qualitative agreement with our experiments. Does a model need to accurately account for these changes in the width of relaxation time distribution in order to predict the real mechanical properties of polymer glasses? Or are these changes in the width of distribution a "detail" with little consequence for mechanical properties?

There is an opportunity to unify the description of the deformation of different types of glassy materials. While shear transformation zones feature prominently in the metallic glass literature they are rarely mentioned in the description of polymer glass deformation. We expect that shear transformation zones are equally important for different classes of glassy materials. Simulations can play a critical role in testing theoretical approaches on different classes of glassy materials; we anticipate that some unification of the theoretical approaches taken in these two communities should be possible.

*Acknowledgements*

MDE gratefully acknowledges extensive and useful collaboration with Juan de Pablo, Rob Riggleman, Jim Caruthers, Grisha Medvedev, Ken Schweizer, Hau-Nan Lee, Steve

Swallen, Keewook Paeng, Ben Bending, and Josh Ricci. In particular, MDE thanks Jim Caruthers for providing the ideas that led to our work on polymer glass deformation. We gratefully acknowledge the support of the National Science Foundation, Division of Materials Research, Polymers Program for support of this work (1404614, 1104770, 0907607).

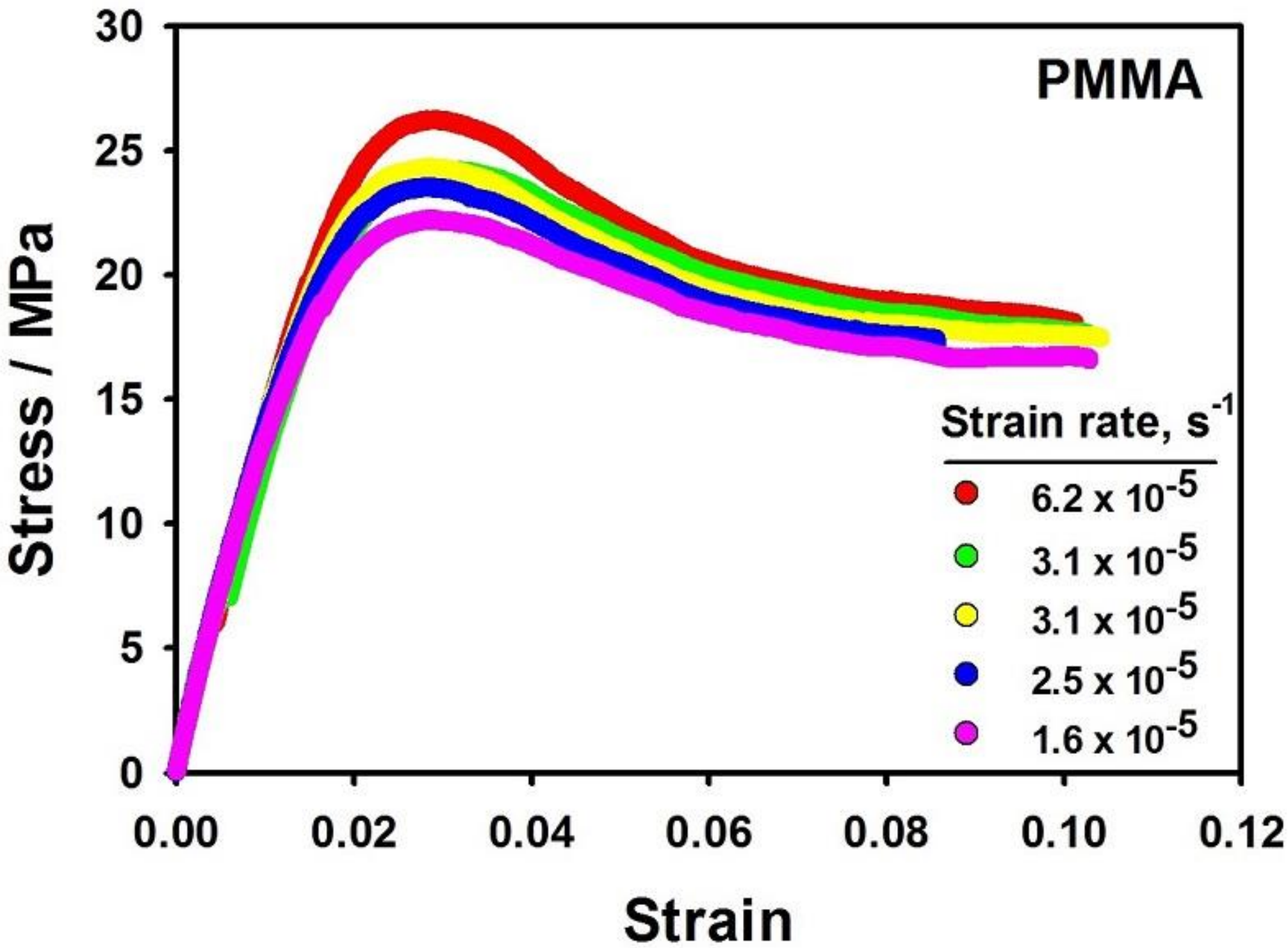

Figure 1. Mechanical response of PMMA deformed at constant strain rates as listed in the legend. Engineering stress (extension force divided by cross-sectional area) is plotted against global strain (fraction increase in sample length). These tests were performed at $T_g$-19 K ($T_g$ = 392 K). Data for the four highest strain rates from reference 1.

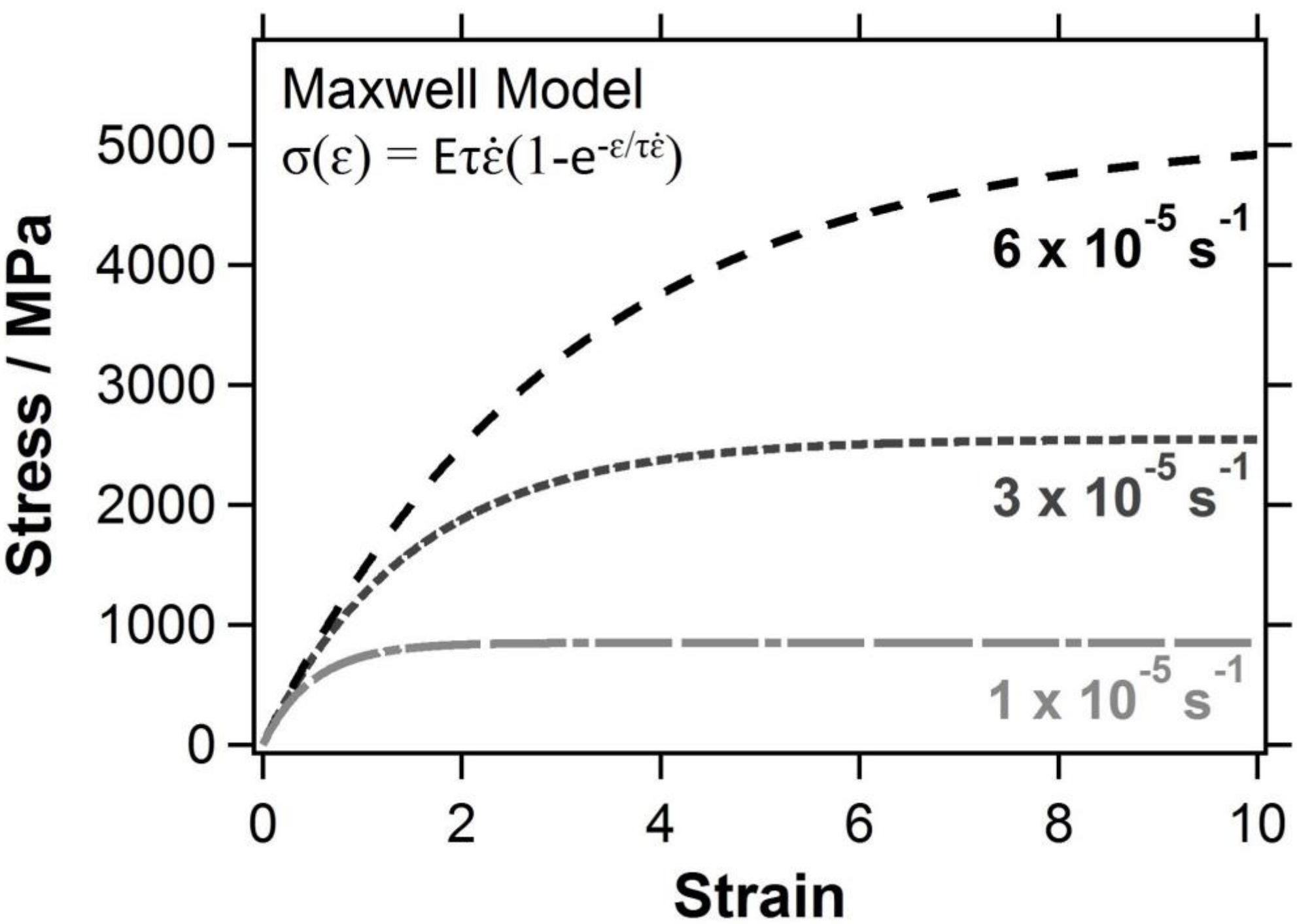

Figure 2. Maxwell model calculations for three constant strain rate deformations using parameters for the PMMA glass of Figure 1. After an initial linear regime at small strains, individual curves reach a steady state flow stress, $\sigma_{flow}$, at higher strains. For the Maxwell model, the flow stress depends linearly on the strain rate in contrast to the data shown in Figure 1.

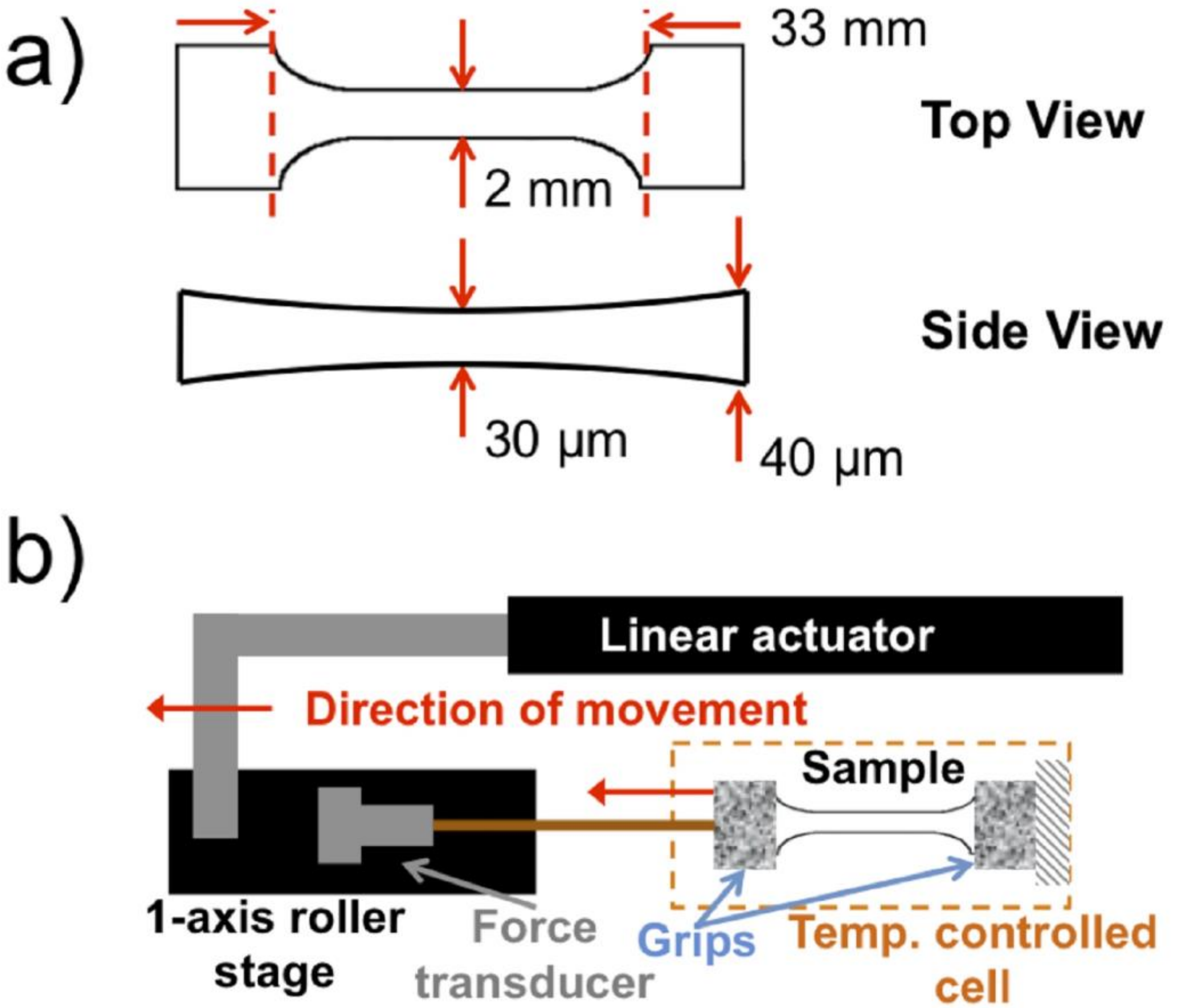

Figure 3. Sample geometry and schematic of deformation apparatus. Panel a shows the dimensions of a typical sample prior to deformation. Panel b is a schematic of the deformation apparatus that sits atop a confocal optical microscope; a top view is shown. A programmable linear actuator drives the deformation. Reprinted with permission from Bending et al. *Macromolecules* **2014,** *47*, 800-806. Copyright 2014 American Chemical Society.

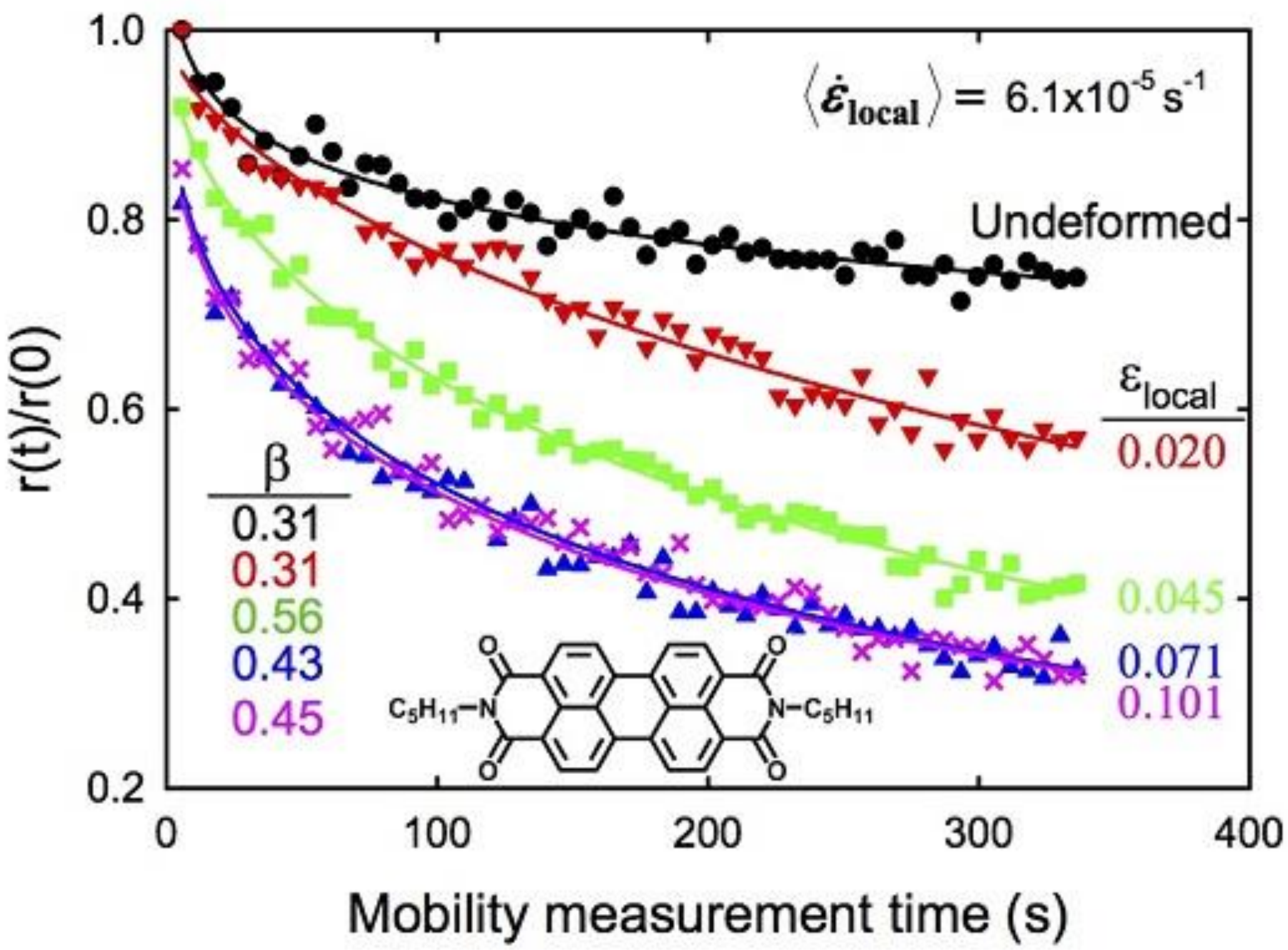

Figure 4. Anisotropy decay data at various strains during constant strain rate deformation of a PMMA glass at $T_g$-19 K. Local strains and the value of the KWW β parameter are shown in the legends. Solid lines through the data are fits to the KWW function. The structure of the probe DPPC is shown. Reprinted with permission from Bending et al. *Macromolecules* **2014,** *47*, 800-806. Copyright 2014 American Chemical Society.

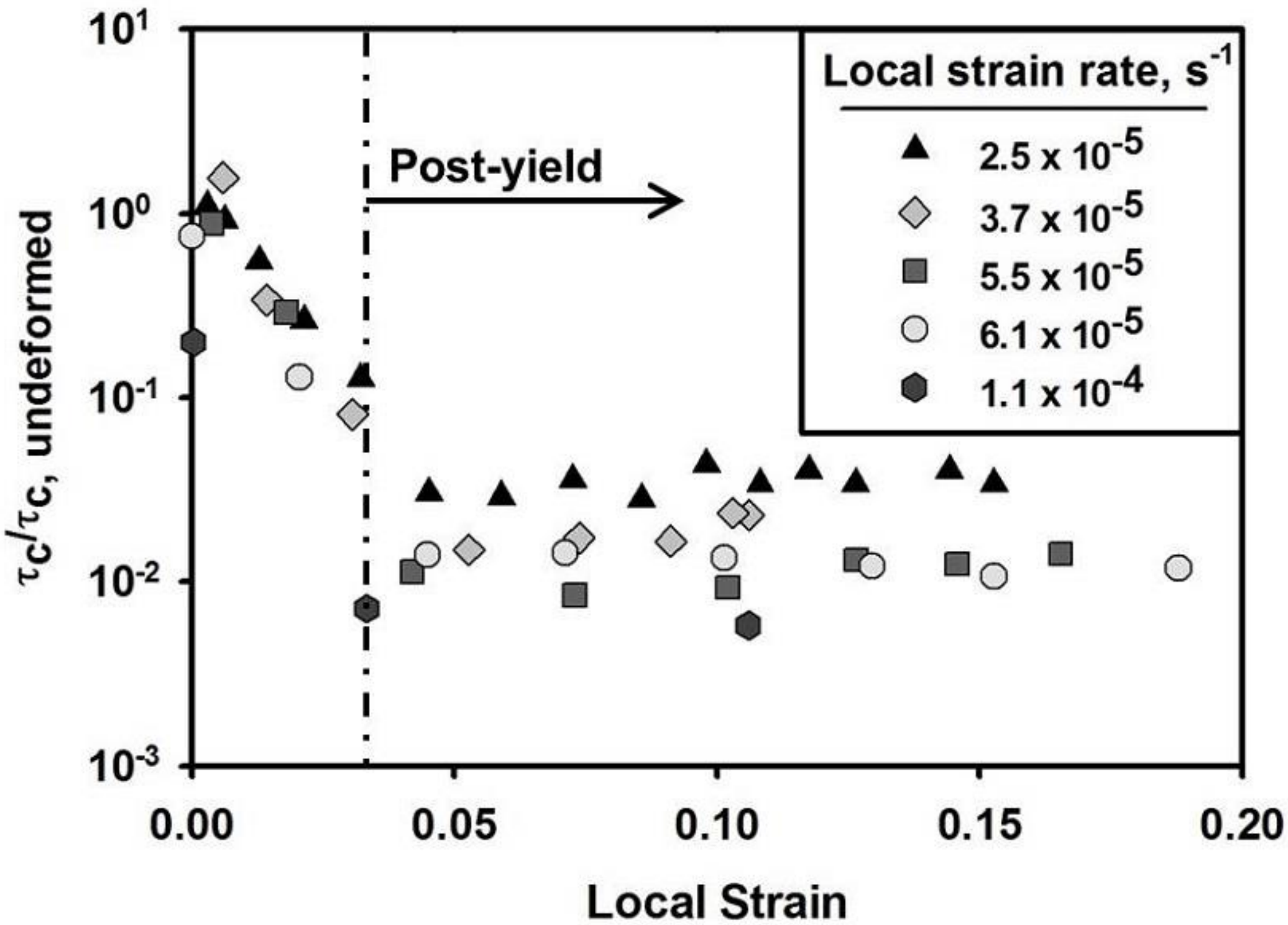

Figure 5. Evolution of segmental mobility with increasing local strain during constant strain rate deformation of a PMMA glass at $T_g$-19 K. Average post-yield local strain rates are displayed in the legend. The y-axis scales $\tau_c$ during deformation to its value for an undeformed glass with the same thermal history. Data for the four highest local strain rates were reported in reference 1.

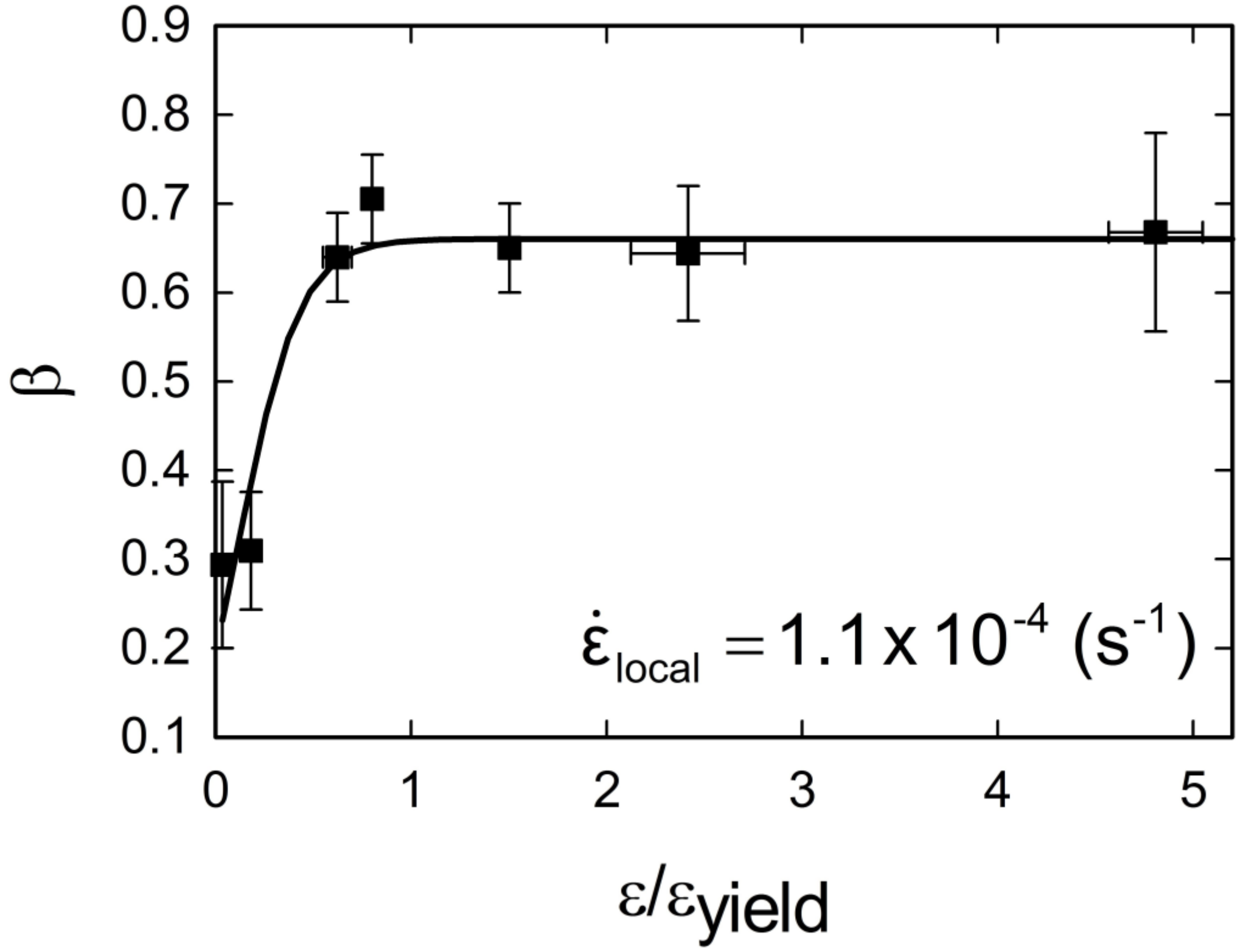

Figure 6. Evolution of the KWW β parameter during constant strain rate deformation of a PMMA glass deformed at $T_g$-19 K. The x-axis scales local strain by the strain at yield. The KWW β parameter increases from its pre-deformation value of ~0.31 up until yield. After yield, β levels off. The average post-yield local strain rate is displayed in the legend. Guides to the eye are shown as solid lines.

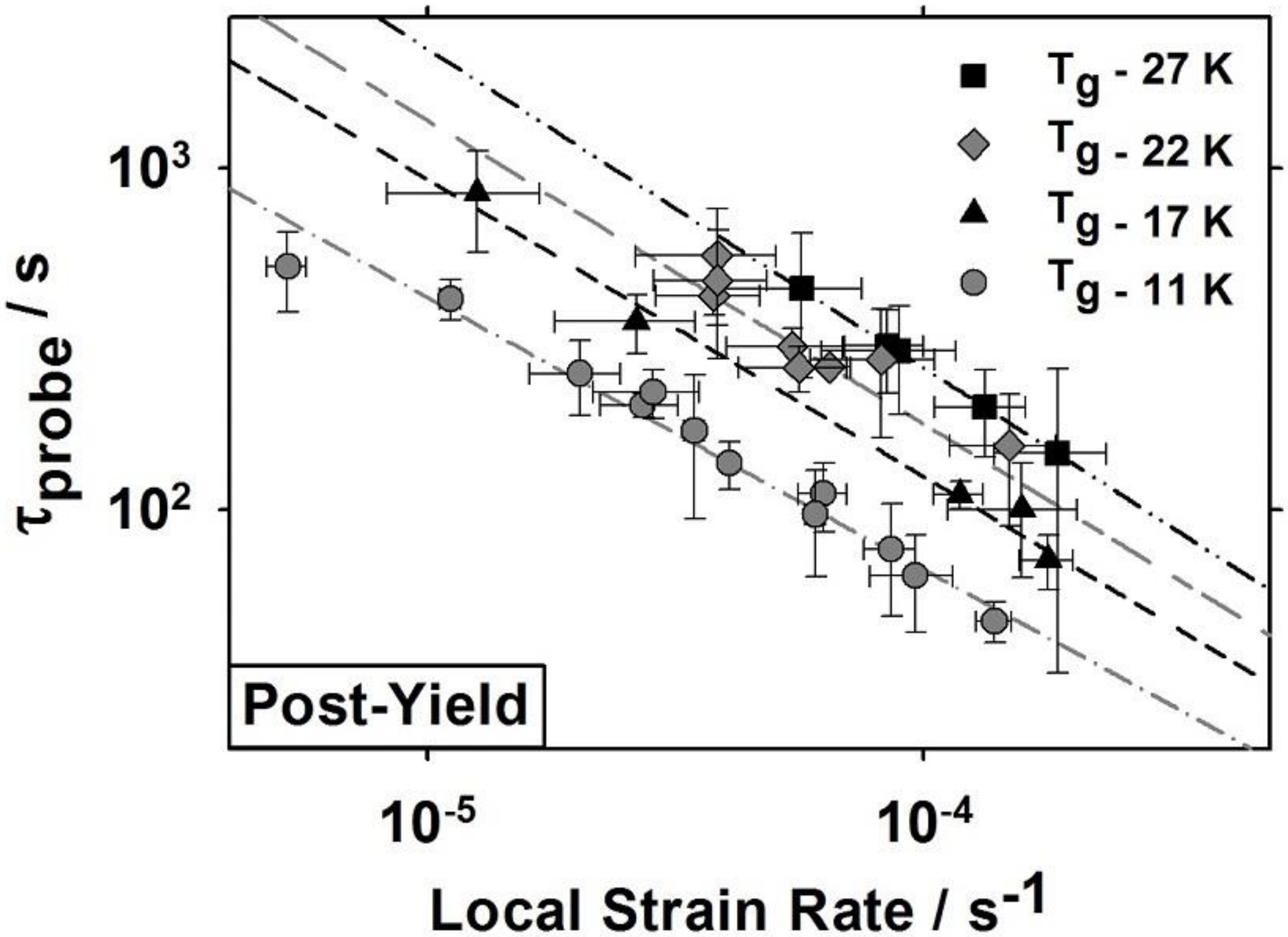

Figure 7. Dependence of average post-yield $\tau_{probe}$ values on local strain rate for PMMA glasses at four different temperatures undergoing constant strain rate deformation. Within each temperature series, $\tau_{probe}$ decreases with increasing strain rate. At a fixed strain rate, $\tau_{probe}$ decreases with increasing temperature. Each data point corresponds to the average post-yield data for one deformation; error bars represent one standard deviation in the data. Dashed lines represent power law fits to the data. Data from reference 39.

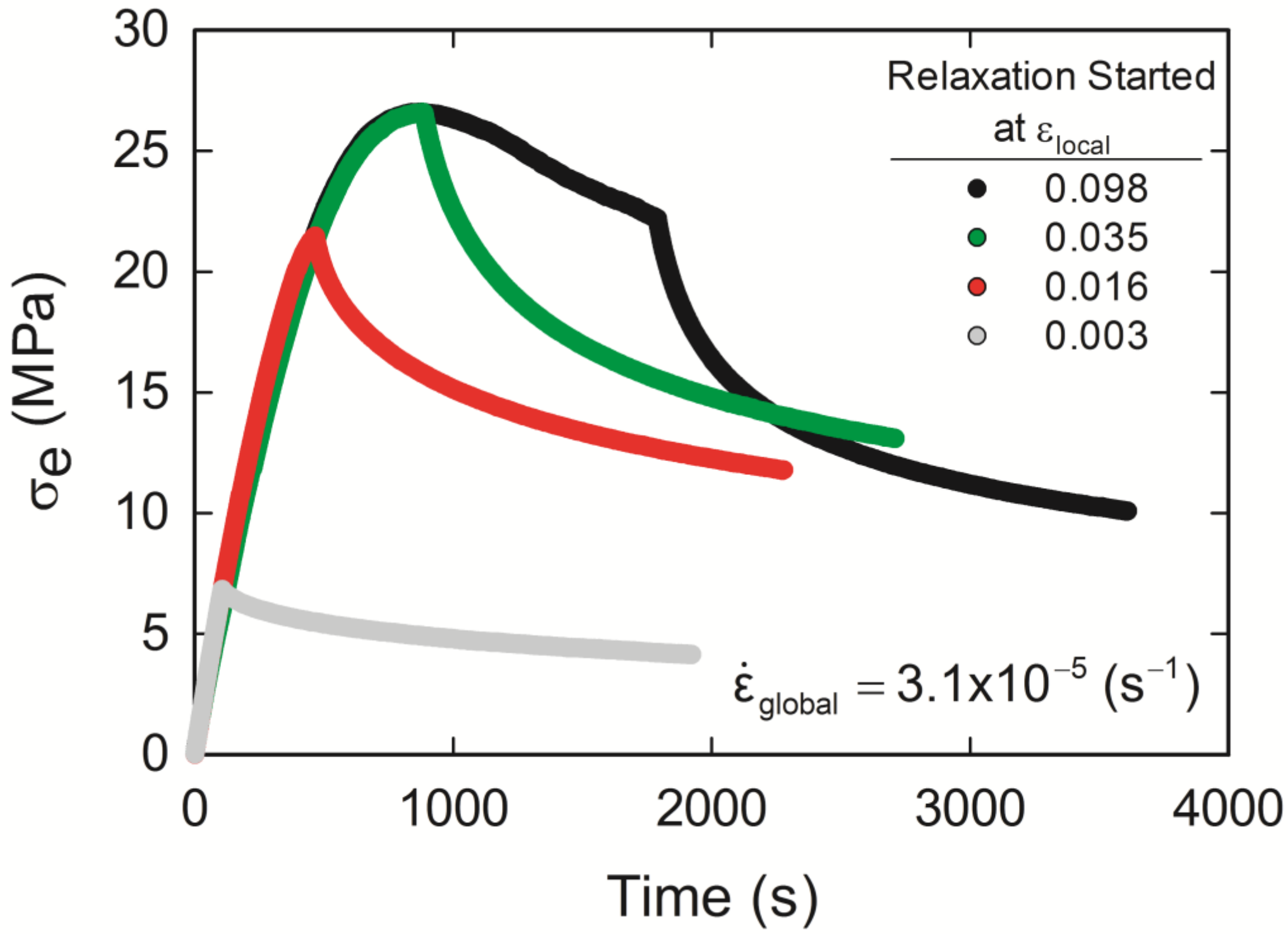

Figure 8. Stress as a function of time for a PMMA glass at $T_g$-19 K deformed at a constant strain rate of $3.1 \times 10^{-5}$ $s^{-1}$, followed by stress relaxation at various strains, as displayed in the legend. For clarity, only part of the stress-relaxation data is shown for each test.

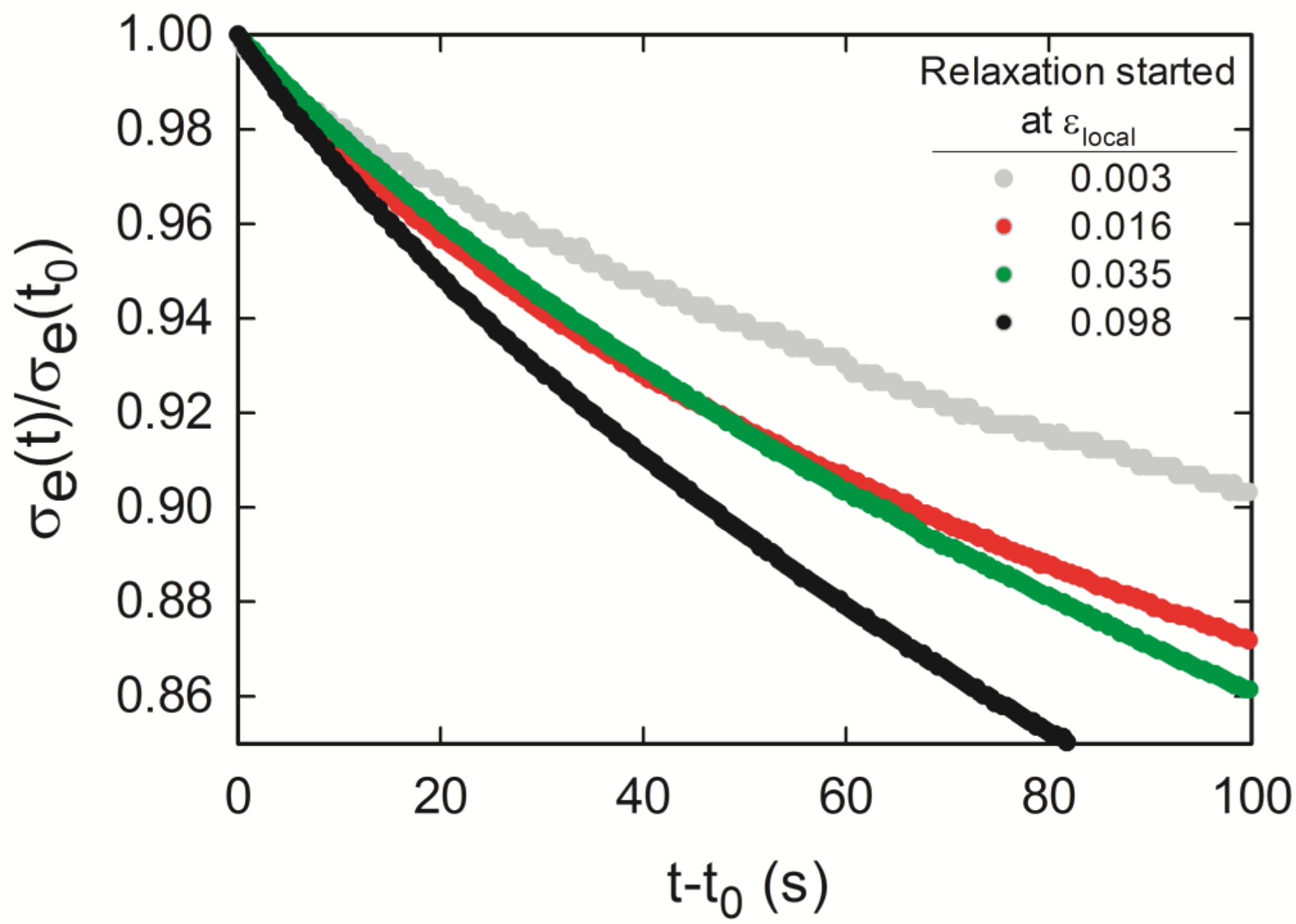

Figure 9. Stress relaxation response for the data in Figure 8 normalized to the initial stress and shifted to the starting time of stress relaxation. The local strain at which stress relaxation was initiated is provided in the legend. For clarity, only the first 100 seconds of stress relaxation data is shown.

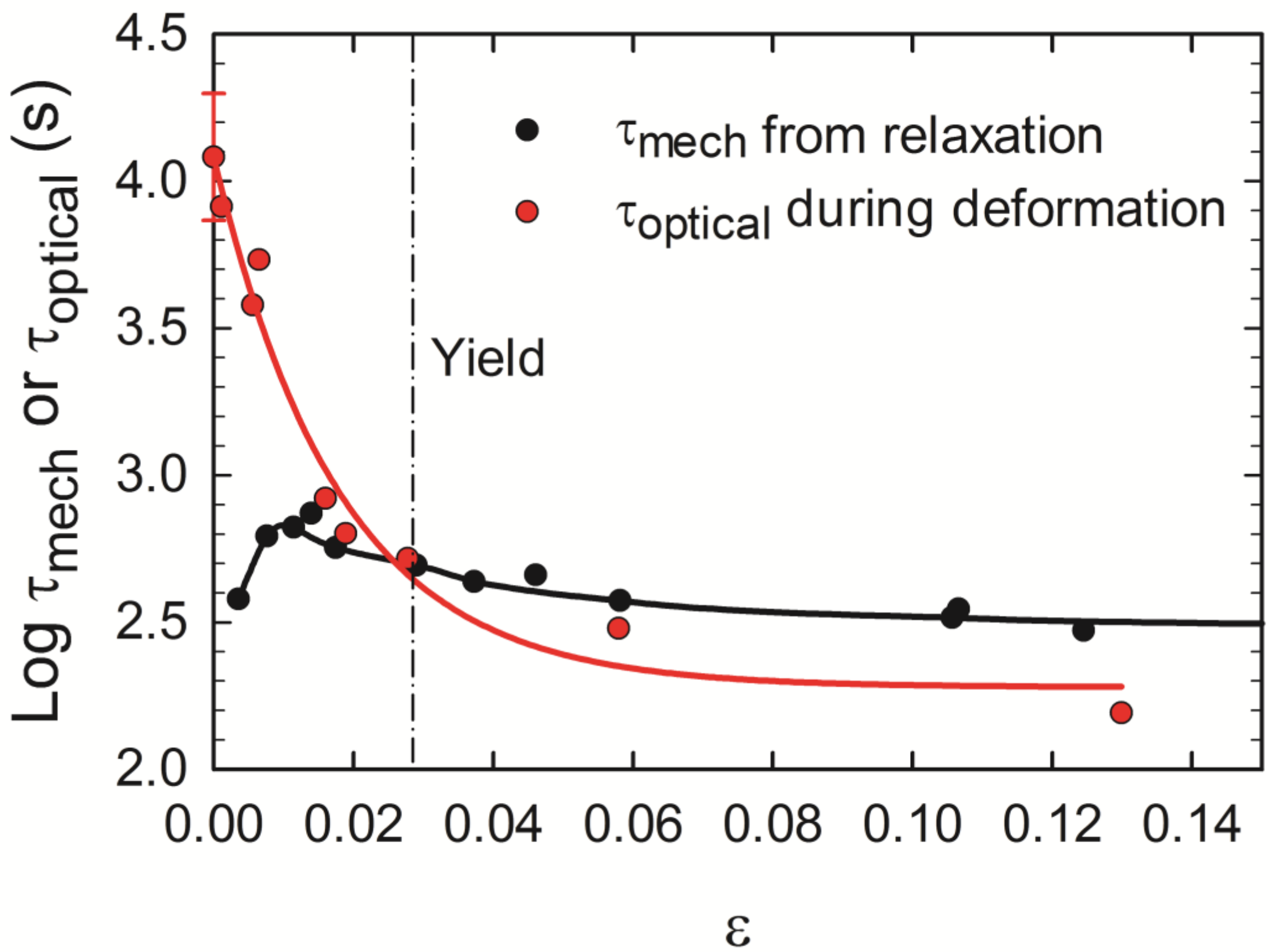

Figure 10. Evolution of $\tau_{mech}$, as determined by stress relaxation, and optically-measured $\tau_{probe}$ with strain during the constant strain rate deformations of Figures 8 and 9. During deformation, $\tau_{probe}$ experiences almost a 100-fold decrease, which is not observed in $\tau_{mech}$. The x-axis scales the local strain to the strain at yield. Solid lines are guides to the eye.

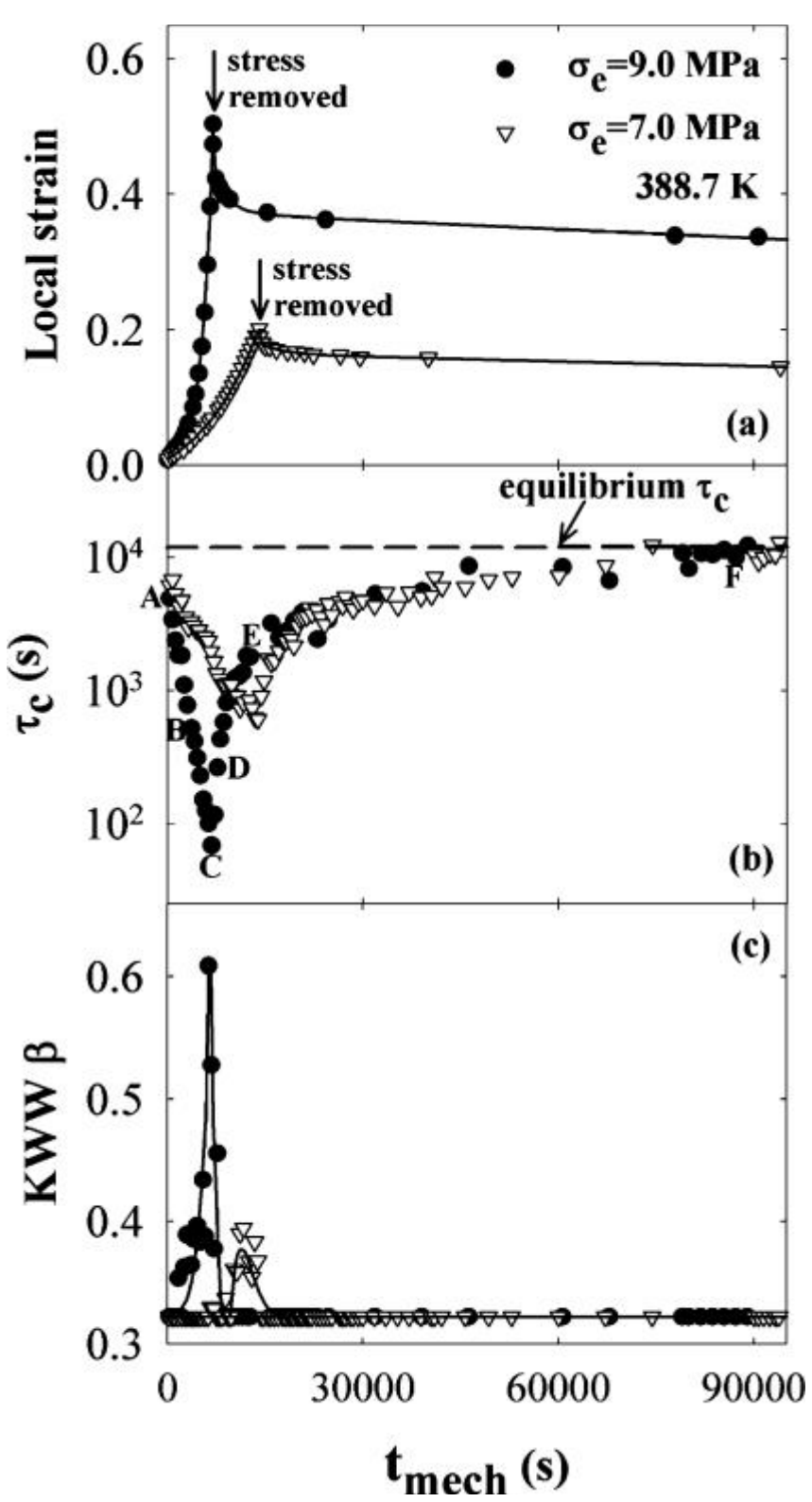

Figure 11. Local strain, $\tau_c$, and the KWW β parameter during creep and recovery of PMMA aged to equilibrium at $T_g$-6 K. Engineering stress in each of the two deformations is shown in the legend. In Panel a, it is observed that strain is not fully recovered after the deformation, despite the recovery of the equilibrium $\tau_c$ (panel b). Panel b shows enhancement of segmental mobility by up to a factor of ~100 during deformation, and the evolution of $\tau_c$ into equilibrium after deformation. During deformation, the KWW β parameter (panel c) drastically increases while stress is applied but rapidly recovers its pre-deformation value after stress is released. Reprinted with permission from Lee and Ediger, *Macromolecules* **2010,** *43*, 5863-5873. Copyright 2010 American Chemical Society.

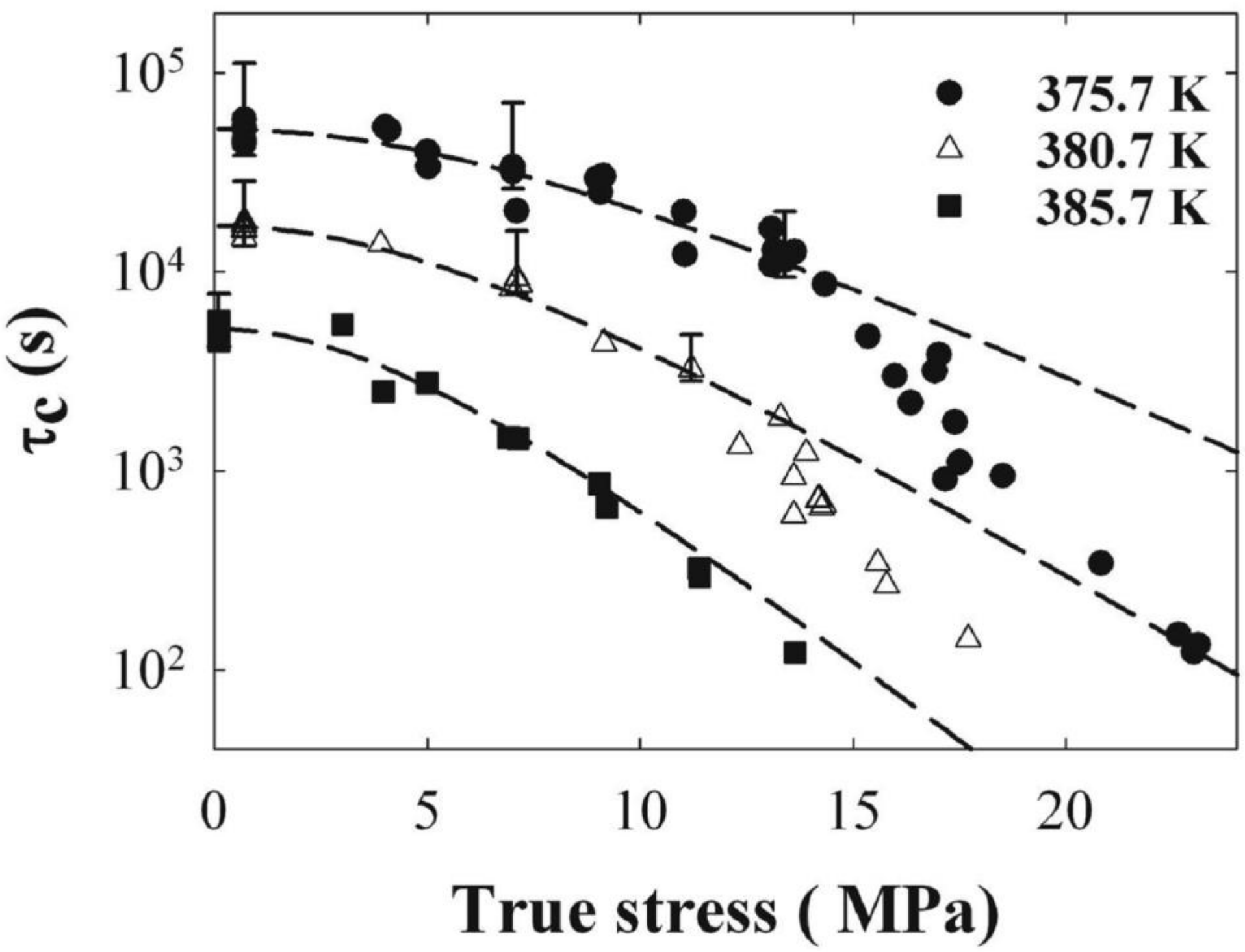

Figure 12. $\tau_c$ as a function of true stress during constant stress deformation of PMMA glasses at three temperatures, as shown in the legend. At low stress, the data agrees with the Eyring model (dashed lines). $\tau_c$ deviates from the Eyring model during flow at higher stresses. Reprinted with permission from Lee et al., *J. Polym. Sci. B Polym. Phys.* **2009,** *47*, 1713-1727. Copyright 2009 Wiley Periodicals, Inc.